\documentclass[english,notitlepage,nofootinbib]{revtex4-1}
\usepackage{amsmath}
\usepackage{graphicx}
\usepackage{amssymb}
\usepackage{float}
\usepackage{mathpazo}
\usepackage{lineno}
\usepackage{subfigure}
\usepackage[utf8]{inputenc}
\usepackage{color}
\usepackage[normalem]{ulem}

\newcommand{\ignore}[1]{}

\newcommand{\be}{\begin{equation}}
\newcommand{\ee}{\end{equation}}
\def \ba#1\ea{\begin{align}#1\end{align}}
\newcommand{\bit}{\begin{itemize}}
\newcommand{\eit}{\end{itemize}}

\def \slashb#1{\setbox0=\hbox{$#1$}#1\hskip-\wd0\dimen0=5pt\advance
        \dimen0 by-\ht0\advance \dimen0 by\dp0\lower0.5\dimen0\hbox
          to\wd0{\hss \sl/\/ \hss}}

\input{epsf}
\begin{document}
	
\title{Borexino and General Neutrino Interactions}

\author{Amir N.\ Khan $^{a, b}$}   \email{amir.khan@mpi-hd.mpg.de, akhan@fnal.gov}
\author{Werner Rodejohann $^{a}$} \email{werner.rodejohann@mi-hd.mpg.de}
\author{Xun-Jie Xu $^{a}$}   \email{xunjie.xu@mpi-hd.mpg.de}
\affiliation{
$^{a}$Max-Planck-Institut f\"ur Kernphysik, Postfach
103980, D-69029 Heidelberg, Germany\\
$^{b}$Theoretical Physics Department, Fermi National Accelerator Laboratory, P.O.
Box 500, Batavia, IL 60510, USA}

\begin{abstract}
\noindent We derive constraints on all possible general
neutrino-electron interactions (scalar, vector, pseudo-scalar, axial-vector
and tensor) using the recent real time Borexino event rate measurements of $pp$, $pep$ and $^{7}$Be solar neutrinos. Some of the limits improve from TEXONO and CHARM-II for incoming
electron and muon neutrinos while the rest remains weaker for Borexino and those for the tau flavor are the first ones.  Future improvements by next-generation solar
neutrino experiments are also studied. The limits extend the physics reach of solar neutrino measurements to  TeV-scale physics. 
Finally, the different properties of the new interactions for Dirac and Majorana neutrinos are 
discussed. 
\end{abstract}

\maketitle

\section{Introduction}

\noindent Precision studies in neutrino physics allow to tighten the
parameters of the standard three neutrino paradigm \cite%
{Tanabashi:2018oca}. Indeed, the precision on the parameters of the PMNS
matrix approaches the one of the corresponding CKM parameters, see e.g.\ 
\cite{deSalas:2018bym}. Moreover, effects of new physics beyond the standard
paradigm can be tested.

In this paper we will analyze the possible presence of neutrino-electron
interactions beyond the usual $V-A$ structure within the Standard Model.
Since neutrinos are often considered a prime window to new physics, it is
natural to assume such new interactions for neutrinos.
Taking a general parametrization originally studied in Refs.~\cite%
{Rosen:1982pj,Shrock:1981cq,Kayser:1981nw}, which
considers all Lorentz-invariant possibilities (scalar, vector, pseudoscalar,
axialvector and tensor)\footnote{It is interesting to point out that since Majorana neutrinos do not have flavor-diagonal vector and tensor interactions, a study of electron-neutrino cross sections, in principle, allows to distinguish Dirac from Majorana neutrinos \cite{Rosen:1982pj,Kayser:1981nw,Rodejohann:2017vup}. We will expand more discussions on this issue  in Sec.~\ref{sec:diffDM}.
}
 in neutrino-electron scattering, we exploit the Borexino solar neutrino measurements \cite{Bellini:2011rx,Collaboration:2011nga,Bellini:2014uqa,Bellini:2013lnn,Borexino:2017fbd}
 to set limits on the size of the new interactions. While new
vector interactions are quite often studied, known as non-standard
interactions \cite{Davidson:2003ha,Ohlsson:2012kf,Farzan:2017xzy,Esteban:2018ppq}, different Lorentz structures are largely
unexplored (neutrino interactions with a structure different from vector do
not lead to observable matter effects in neutrino oscillations \cite%
{Bergmann:1999rz}). Some existing studies can be found in 
\cite{Healey:2013vka,Sevda:2016otj,Lindner:2016wff,Heurtier:2016otg,Rodejohann:2017vup,Kosmas:2017tsq,Yang:2018yvk,AristizabalSierra:2018eqm,Bischer:2018zcz,Blaut:2018fis,Ge:2018uhz,Bolton:2019wta,Link:2019pbm,Bischer:2019ttk}.

We will employ the Borexino measurements of low energy $pp$, $pep$ and $^{7}$Be solar neutrinos. 
As the originally produced electron neutrinos
oscillate to muon and tau neutrinos, this allows to set limits on
general interactions of all flavors. Previously, limits 
on general neutrino-electron interactions 
were obtained in
Ref.\  \cite{Rodejohann:2017vup} using TEXONO \cite{tex} and CHARM-II \cite%
{cii} data for electron and muon neutrinos, respectively. We will improve
several of those limits, and set the very first ones on general tau
neutrino interactions. Possible future limits by upcoming of hypothetical
solar neutrino measurements are also estimated. If percent-level coupling strengths are measured, and the new interactions  are interpreted in terms of new exchanged bosons, then
new physics of weak and TeV scales is tested by solar neutrino experiments. 
\newline


The paper is structured as follows. In Sec.\  \ref{sec:form} we set up the
formalism of neutrino-electron scattering with general interactions.
Section \ref{sec:fit} describes the data and fit procedure that we follow,
with the results being  discussed in Sec.\  \ref{sec:res}. 
We also address the differences between Dirac
and Majorana neutrinos in this framework in Sec.\  \ref{sec:diffDM}, conclusions
are drawn in Sec.\  \ref{sec:concl}.

\section{ \label{sec:form} Neutrino-electron scattering in the presence of
general neutrino interactions}

\noindent In this section we lay out the general formalism to describe
general neutrino interactions relevant for elastic
neutrino-electron scattering. Starting with the Standard Model (SM), we have
neutral current (NC) and charged current (CC) interactions between the
target electrons and the three flavors of neutrinos. To be specific, in the SM, $\nu
_{e}e$-scattering involves the CC and NC interactions, while $\nu _{\mu /\tau
}e$-scattering depends only on NC interactions. 
The effective SM Lagrangian for the NC interactions is given as
\begin{equation}
\mathcal{L}_{\mathrm{NC}}^{\mathrm{SM}}=\frac{G_{F}}{\sqrt{2}}\left[ 
\overline{\nu }\gamma ^{\mu }(1-\gamma ^{5})\nu \right] \left[ \overline{%
\ell }\gamma _{\mu }(g_{V}^{\ell }-g_{A}^{\ell }\gamma ^{5})\ell \right] ,
\label{eq:LSMNC}
\end{equation}%
where the vector and axial vector couplings are 
\begin{equation}
g_{V}^{\ell }=-\frac{1}{2}+2\sin ^{2}\theta _{W}\text{ and }g_{A}^{\ell }=-%
\frac{1}{2}\,.  \label{eq:LSM1}
\end{equation}%
For the CC interactions, after a Fierz transformation one can write (flavor indices are suppressed)
\begin{equation}
\mathcal{L}_{\mathrm{CC}}^{\mathrm{SM}}=\frac{G_{F}}{\sqrt{2}}\left[ 
\overline{\nu }\gamma ^{\mu }(1-\gamma ^{5})\nu \right] \left[ \overline{%
\ell }\gamma _{\mu }(1-\gamma ^{5}))\ell \right] .  \label{eq:LSMCC}
\end{equation}
We are interested here in new neutrino physics that may show up in the form
of general neutrino interactions. With this we denote  new
interactions for neutrino-electron scattering, that can be scalar,
pseudoscalar, vector, axialvector or tensor. The effective four-fermion
interaction Lagrangian is 
\begin{equation}
\Delta \mathcal{L}=\frac{G_{F}}{\sqrt{2}}\sum_{\substack{ a=S,P,  \\ V,A,T}}(%
\overline{\nu _{\alpha }}\Gamma ^{a}\nu _{\beta })\left[ \overline{\ell }%
\Gamma ^{a}(\epsilon _{\alpha \beta }^{a}+\tilde{\epsilon}_{\alpha \beta
}^{a}{\,i^{a}}\gamma ^{5})\ell \right] ,  \label{eq:LBSM}
\end{equation}%
where $\Gamma ^{a}\equiv \left \{ I,{i}\gamma ^{5},\  \gamma ^{\mu },\  \gamma
^{\mu }\gamma ^{5},\sigma ^{\mu \nu }\equiv \frac{i}{2}[\gamma ^{\mu
},\gamma ^{\nu }]\right \} $ are the five fermion operators, corresponding
to Scalar ($S$), Pseudoscalar ($P$), Vector ($V$), Axialvector ($A$) and
Tensor ($T$), respectively. Furthermore, following the convention in Ref.\  \cite{Rodejohann:2017vup}, we have $i^{a}=i$ for $a=(S,\ P,\ T)$%
 and $i^{a}=1$ for $a=(V,\ A)$.
Including the factor $i$ for the $S,P,T$ interactions is necessary to have $\epsilon _{\alpha
\alpha }$ and $\tilde{\epsilon}_{\alpha \alpha }$ real. We assume that $%
\epsilon $ and $\tilde{\epsilon}$ are hermitian matrices, i.e., $\epsilon
_{\alpha \beta }=\epsilon _{\beta \alpha }^{\ast }$ and $\tilde{\epsilon}%
_{\alpha \beta }=\tilde{\epsilon}_{\beta \alpha }^{\ast }$, so that Eq.~(\ref%
{eq:LBSM}) is self-conjugate\footnote{%
Otherwise the hermitian conjugate ($h.c.$) term should be added, which again
leads to hermitian $\epsilon $ and $\tilde{\epsilon}$\label{fn:hermitian}.}.
Possible phases of the $\epsilon $ and $\tilde{\epsilon}$ matrices are
ignored in what follows.

For Majorana neutrinos some of the interactions in Eq.~(\ref{eq:LBSM})
cannot be written in terms of Majorana spinors. More specifically, in this
case, the vector and tensor interactions with $\alpha=\beta$ for each
generation should vanish (i.e., $\epsilon^{V}_{\alpha \alpha}=\tilde{\epsilon%
}^{V}_{\alpha \alpha}=\epsilon^{T}_{\alpha \alpha}=\tilde{\epsilon}%
^{T}_{\alpha \alpha}=0$), which is a known property of Majorana spinors \cite%
{Rosen:1982pj, Dass:1984qc, Bergmann:1999rz}. However, considering three
generations of neutrinos, such interactions can still exist for $\alpha \neq
\beta$. Nevertheless, for Majorana neutrinos, the parameter space of the
general Lorentz-invariant interactions is smaller than the one for Dirac
neutrinos. Our analyses will be focused on Dirac neutrinos, we will address
the difference to the Majorana case  in Sec.~\ref{sec:diffDM}%
. Furthermore, we focus on flavor diagonal interactions, i.e.\ we constrain $%
\epsilon_{\alpha \alpha}^a$ and $\tilde \epsilon_{\alpha \alpha}^a$.

The limits on the off-diagonal terms will be very similar to the diagonal ones for scalar, pseudo-scalar and tensor interactions due to the absence of interference with the SM. For vector or axial-vector interactions, this case has been well-studied in the context of the usual NSI in Ref.\ \cite{Khan:2017oxw} for the Borexino data.

In general the new interactions of Eq.\ (\ref{eq:LBSM}) are added to the SM
interactions in Eqs.\ (\ref{eq:LSMNC}) and (\ref{eq:LSMCC}). The differential
cross section of neutrino-electron scattering is found to be \cite{Rodejohann:2017vup}:
\begin{equation}
\frac{d\sigma }{dT}(\nu_{\alpha}+e^{-}\rightarrow \nu_{\beta}+e^{-})=\frac{%
G_{F}^{2}m_{e}}{2\pi }\left[ A_{\alpha \beta}+2B_{\alpha \beta}\left(1-\frac{%
T}{E_{\nu }}\right)+C_{\alpha \beta}\left(1-\frac{T}{E_{\nu }}%
\right)^{2}+D_{\alpha \beta} \frac{m_{e}T}{4E_{\nu }^{2}}\right],
\label{diffcros}
\end{equation}%
where $m_{e}$ is the electron mass, $E_{\nu }$ is the neutrino energy and $T$
is the electron recoil energy. The parameters $A_{\alpha \beta}$, $B_{\alpha
\beta}$, $C_{\alpha \beta}$ and $D_{\alpha \beta}$ are defined as (given here for complex parameters and ignoring flavor indices for simplicity) 
\begin{eqnarray}
A & = & \frac{1}{4}\left|\epsilon^{A}+\epsilon^{V}-\tilde{\epsilon}^{A}-\tilde{\epsilon}^{V}+2g^{L}\right|^{2}+\frac{1}{8}\left|\epsilon^{S}+i\tilde{\epsilon}^{P}\right|^{2}+\frac{1}{8}\left|\epsilon^{P}+i\tilde{\epsilon}^{S}\right|^{2}\nonumber \\
 &  & +\left|\epsilon^{T}-i\tilde{\epsilon}^{T}\right|^{2}+\frac{1}{2}{\rm Re}\left[\left(\epsilon^{T}-i\tilde{\epsilon}^{T}\right)^{*}\left(\epsilon^{P}+i\tilde{\epsilon}^{S}-\epsilon^{S}-i\tilde{\epsilon}^{P}\right)\right],\nonumber \\
B & = & -\frac{1}{8}\left|\epsilon^{P}+i\tilde{\epsilon}^{S}\right|^{2}-\frac{1}{8}\left|\epsilon^{S}+i\tilde{\epsilon}^{P}\right|^{2}+\left|\epsilon^{T}-i\tilde{\epsilon}^{T}\right|^{2},\nonumber \\
C & = & \frac{1}{4}\left|\epsilon^{A}+\tilde{\epsilon}^{A}-\epsilon^{V}-\tilde{\epsilon}^{V}-2g^{R}\right|^{2}+\frac{1}{8}\left|\epsilon^{S}+i\tilde{\epsilon}^{P}\right|^{2}+\frac{1}{8}\left|\epsilon^{P}+i\tilde{\epsilon}^{S}\right|^{2}  \label{ABCD-val} \\
 &  & +\left|\epsilon^{T}-i\tilde{\epsilon}^{T}\right|^{2}-\frac{1}{2}{\rm Re}\left[\left(\epsilon^{T}-i\tilde{\epsilon}^{T}\right)^{*}\left(\epsilon^{P}+i\tilde{\epsilon}^{S}-\epsilon^{S}-i\tilde{\epsilon}^{P}\right)\right],\nonumber \\
D & = & {\rm Re}\left[\left(\epsilon^{A}+\epsilon^{V}-\tilde{\epsilon}^{A}-\tilde{\epsilon}^{V}+2g^{L}\right)\left(\epsilon^{A}+\tilde{\epsilon}^{A}-\epsilon^{V}-\tilde{\epsilon}^{V}-2g^{R}\right)^{*}\right]\nonumber \\
 &  & -4\left|\epsilon^{T}-i\tilde{\epsilon}^{T}\right|^{2}+\left|\epsilon^{S}+i\tilde{\epsilon}^{P}\right|^{2}. \nonumber \label{ABCD-val}
\end{eqnarray}
To recover the explicit flavor indices, one only needs to add
subscripts ${}_{\alpha \beta}$ to all quantities in Eq.~(\ref{ABCD-val}); in
addition one has 
\begin{equation}
\left(g^{L}_{\alpha \beta},\ g^{R}_{\alpha \beta}\right)=%
\begin{cases}
\left(2\sin^{2}\theta_{W}+1,\ 2\sin^{2}\theta_{W}\right) & 
(\alpha=\beta=e)\,, \\ 
\left(2\sin^{2}\theta_{W}-1,\ 2\sin^{2}\theta_{W}\right) & 
(\alpha=\beta=\mu,\  \tau)\,, \\ 
0 & (\alpha \neq \beta)\,.%
\end{cases}
\label{eq:g1g2}
\end{equation}
Note that the SM couplings appear only in $A$, $C$ and $D$. The term
proportional to $B$ is a pure new physics term that contains no SM
contribution. 

We restrict our analysis to the total event rates.  
In this case the total cross sections in terms of the maximum recoil
energy of electrons $T^{\max }(E_{\nu })$ are the relevant observable.
We can obtain the total cross section from Eq.\ (\ref{diffcros})
as 
\begin{equation}
\sigma (\nu e)=\frac{G_{F}^{2}m_{e}T_{\max }}{2\pi }\left[ A+2B\left( 1-%
\frac{T_{\max }}{2E_{\nu }}\right) +C\left( 1+\frac{T_{\max }^{2}}{3E_{\nu
}^{2}}-\frac{T_{\max }}{E_{\nu }}\right) +D\frac{m_{e}T_{\max }}{8E_{\nu
}^{2}}\right] ,  \label{eq:xsABCD}
\end{equation}%
where $T_{\max }(E_{\nu })\equiv E_{\nu }/(1+m_{e}/2E_{\nu })$. Here we have 
$E_{\nu }<0.420$ MeV for the continuous \textit{pp} spectrum,
and $E_{\nu }$= 0.862 MeV (1.44 MeV) for neutrinos from $^{7}$Be (\textit{pep}) reactions,
respectively. For each spectrum we have reproduced with good agreement the
expected event numbers quoted by Borexino in \cite{Bellini:2014uqa}. More
details will be discussed in Sec.\  \ref{sec:ana}. It is important to note that the term proportional to 
$C$ in Eqs.\ (\ref{diffcros}) and (\ref{eq:xsABCD}) is suppressed by the kinematic factor 
proportional $\frac{T}{E_{\nu }}$ with respect to $A$. This naturally leads to a relatively tighter 
constraints on the parameters related to $C$. For anti-neutrinos $A$ and $C$ are replaced 
with each other in the cross sections. 

As stated earlier, the cross sections in principle contain contributions
both from flavor conserving and flavor violating processes. For simplicity,
we will restrict ourselves to the flavor conserving case at the neutrino
vertex, i.e.\ $\nu_e \,e \to \nu_e \,e$ and $\nu_{\mu, \tau} \,e \to
\nu_{\mu, \tau} \,e$ scattering. As a consequence there are interference
terms for the SM and new physics terms in the cross sections in Eqs.\ (\ref%
{diffcros}) and (\ref{eq:xsABCD}). Regarding those interference terms, note that
there is no interference of the vector/axial terms with the
scalar/pseudoscalar/tensor-type interaction terms. All such terms cancel out
in the cross amplitude terms due to the products of the odd number of gamma
matrices for vector/axial currents with the even number of gamma matrices
in the scalar/pseudo-scalar/tensor current. Thus, the scalar, pseudo-scalar
and tensor interactions are independent of the vector and axial-vector
currents and in particular do not interfere with the SM interactions. We will discuss this point
in more detail in Sec.\  \ref{sec:diffDM}.




\section{\label{sec:fit} Searching for exotic interactions in solar neutrino
experiments}

\noindent 
In this section we give details of the solar oscillation probabilities,
event rate calculations and the statistical model used for our analysis.

\subsection{Solar Neutrino Oscillation Probabilities}

As solar neutrinos change their flavor from production to
detection, we need to consider the survival probabilities for the $pp$, $%
^{7}Be$ and $pep$ neutrinos that we will use for our model to fit with the
data. We follow the notation from \cite{Khan:2017oxw}. 
If there were no matter effects, the 
oscillation amplitude would be $A_{\alpha \beta }=U_{\alpha i}X_{i}U_{i\beta
}^{\dagger }$, where $i$ are mass indices while $\alpha $ and $\beta $ are
the flavor indices. Summation over the mass indices is implied. Here $U$
is the neutrino mixing matrix and $X$ is the diagonal phase matrix $X\ $=
diag$(1,\exp (-i2\pi L/L_{21}^{osc}),\exp (-i2\pi L/L_{31}^{osc}))$, where
the oscillation length is defined as $L_{ij}^{osc}=4\pi
E_\nu/(m_{i}^{2}-m_{j}^{2})$. Thus, the solar neutrinos oscillation probability
would read 
\begin{equation}
P_{\alpha \beta }=|A_{\alpha \beta }|^{2}=|U_{\alpha i}X_{i}U_{i\beta
}^{\ast }|^{2}\,.  \label{eq:PeeV}
\end{equation}%
Due to the very large distance between Sun and Earth we can take the
averaged oscillation probability as 
\begin{equation}
\langle P\rangle _{\alpha \beta }=U_{\alpha i}U_{\beta i}^{\ast }U_{\alpha
i}^{\ast }U_{i\beta }=|U_{\alpha i}|^{2}|U_{\beta i}|^{2}\,.  \label{eq:Pbar}
\end{equation}%
Expressed in terms of mixing angles, the averaged probability for solar
neutrinos is $\langle P\rangle _{ee}$ = $s_{13}^{4}+(c_{12}c_{13})^{4}$ + $%
(s_{12}c_{13})^{4}$, where $s_{ij}\equiv \sin \theta _{ij}$ and $%
c_{ij}\equiv \cos \theta _{ij}$ in the commonly used notation \cite%
{Tanabashi:2018oca}. 

Matter effects are important for precision studies, and depend on energy.
Solar neutrinos from the low energy $pp$ reaction, which has a continuous
spectrum with energy $E_{\nu }\leq 0.420$ MeV, witness very little matter
effects. The probability $P_{ee}$ has less than a percent difference from
the path-averaged expression in Eq.\ (\ref{eq:Pbar}). However, for the
somewhat higher energy discrete spectra of $^{7}$Be and $pep$ neutrinos ($%
E_{\nu }=0.862$ and $E_{\nu }=1.44$ MeV, respectively), the matter effects
are still small, up to $4$-$5\%$, but not entirely negligible. Therefore, we
include the small modifications due to matter effects according to 
\begin{equation}
\langle P^{m}\rangle _{ee}=s_{13}^{4}+\frac{1}{2}c_{13}^{4}{}(1+\cos 2\theta
_{12}^{m}\cos 2\theta _{12})\,,  \label{eq:Peemat}
\end{equation}%
where%
\begin{equation}
\cos 2\theta _{12}^{m}=\frac{1-N_{e}/N_{e}^{res}}{\sqrt{%
(1-N_{e}/N_{e}^{res})^{2}+\tan ^{2}2\theta _{12}}}
\end{equation}%
is the effective mixing angle inside the Sun, $N_{e}$ is the electron number
density at the center of the Sun, $N_{e}^{res}=\Delta m_{12}^{2}\cos 2\theta
_{12}/(2E_{\nu }\sqrt{2}G_{F})$\ is the electron density in the resonance
region, $\Delta m_{12}^{2}$\ is the solar mass-squared difference, $\theta
_{12}$\ is the solar mixing angle and $G_{F}$\ is the Fermi constant. For
the continuous $pp$ spectrum, we use the electron density at average 
$pp$ production point in the above expressions and assume an exponential
decrease of the density outward from the core in the analytic approximations
as discussed in detail in Ref.\  \cite{lopes}. This is an excellent
approximation for $r>0.1R_{solar}$ \cite{Bahcall:1995mm}.

Taking the current best-fit values of the oscillation parameters \cite%
{Tanabashi:2018oca}, we find the vacuum value $\langle P^{vac}\rangle
_{ee}=0.558$, which is modified to $\langle P^{pp}\rangle _{ee}=0.554$ for $%
pp$, $\langle P^{^{7}{Be}}\rangle _{ee}=0.536$ for $^{7}Be$ and $\langle
P^{pep}\rangle _{ee}=0.529$ for $pep$ neutrinos in the case of  matter effects.

\subsection{Borexino, Event Rate Calculations and the $\protect \chi ^{2}$%
-model\label{sec:ana}}

We will consider five measurements made by the Borexino experiment since 2007
both in phase-I \cite{Bellini:2011rx,Collaboration:2011nga} and phase-II 
\cite{Bellini:2014uqa,Bellini:2013lnn,Borexino:2017fbd} runs. The \textit{pp} spectrum was measured in phase-II only, $^{7}$Be and \textit{pep} spectra
were measured in both phase-I and phase-II with an extensively purified
scintillator in phase-II between December 2011 and May 2016 for total of
1291.51 days. The data obtained from the experiment is given in Table \ref%
{tab:borex}.

\begin{table*}[t]
\begin{center}
\begin{tabular}{l|l|l|l|l}
\hline \hline
flux & event rate (phase-I) & event rate (phase-II) & Our prediction
& \%age error (theo.)  \\ \hline
$pp$ & $-$ & $134\pm 10_{-10}^{+6}$ & $133.5\pm 1.4$ & $1.1\%$ \\ \hline
$^{7}Be$ & $46\pm 1.5_{-1.6}^{+1.5}$ & $48.3\pm 1.1_{-0.7}^{+0.4}$ & $%
48.1\pm 2.8$ & $5.8\%$ \\ \hline
$pep$ & $3.1\pm 0.6\pm 0.3$ & $2.43\pm 0.36_{-0.22}^{+0.15}$ & $2.9\pm 0.04$
& $1.5\%$ \\ \hline \hline
\end{tabular}%
\end{center}
\caption{The measured  event rates in Borexino with statistical
uncertainties ($1 \sigma$) and  the predicted event rates from the standard solar model (SSM). 
Our calculated event rates are given in the last
column. }
\label{tab:borex}
\end{table*}

For all five measurements we take the number of target electrons per 100
tons, $N_{e}^{\rm target}$ = $3.307\times 10^{31}$, as quoted in Ref.\  \cite%
{Borexino:2017fbd}, while taking the $pp$ reaction flux from Ref.\  \cite%
{Bahcall:1995mm}\footnote{%
The details of the \textit{pp} flux calculation have been summarized in the
Appendix of Ref.\ 
\cite{Khan:2017oxw}.}. Since the $^{7}$Be and $pep$ fluxes have discrete
spectra, we treat them as delta functions in our analysis to evaluate the
rate in Eq.\ (\ref{eq:totrate}), see below. In addition, as discovered by all BOREXINO experiments that the data mostly favor the high-metallicity SSM, we use only the high-metallicity fluxes for our analysis to constrain our parameters. The high-metallicity SSM flux values we use are $\phi _{^{7}{%
Be}}=4.48\times 10^{9}$ cm$^{-2}$ s$^{-1}$ at $0.862$ MeV and $\phi
_{pep}=1.44\times 10^{8}$ cm$^{-2}$ s$^{-1}$ at 1.44 MeV in our
calculations. Nevertheless, to investigate the discrepancy between the low-metallicity SSM and the data in the context of a specific model beyond the SM would be more interesting. In order to calculate the expected number of events in the
Borexino detector, we can write down the expression for total rates as 
\begin{equation}
R_{\nu }^{i}=N_{e}^{\rm target}\int_{0}^{E_{max}}dE_{\nu }\phi ^{i}(E_{\nu })\left(
\sigma _{e}(E_{\nu })\langle P^{i}\rangle _{ee}+\sigma _{\mu ,\tau }(E_{\nu
})[1-\langle P^{i}\rangle _{ee}]\right) ,  \label{eq:totrate}
\end{equation}%
where $\langle P^{i}\rangle _{ee}$ is given in Eq.\ (\ref{eq:Peemat}), with
the index \textit{i} indicating whether \textit{pp}, $^{7}$Be or \textit{pep}
neutrinos are considered. The cross sections $\sigma _{e}(E_{\nu })$ and $%
\sigma _{\mu ,\tau }(E_{\nu })$ are given in Eq.\ (\ref{eq:xsABCD}). Using the updated values of Higgs and the top quark masses from PDG we have used the results on the radiative corrections of Ref.\ \cite%
{Bahcall:1995mm}. We find 1.9\% decreasing and 1.1\% increasing quantum corrections for \ $\nu_e \,e \to \nu_e \,e$ and $\nu_{\mu, \tau} \,e \to
\nu_{\mu, \tau} \,e$ scattering, respectively. We have normalized our SM predicted total cross-sections accordingly.  Note that we assume in Eq.\ (\ref{eq:totrate}) equal
fluxes of muon and tau neutrinos, which corresponds to maximal $\theta _{23}$%
. This will imply identical limits for the muon- and tau-flavor $\epsilon $
parameters. For non-maximal $\theta _{23}$ there will in reality be slightly different limits.
For data fitting, we use the
following $\chi ^{2}$-estimator  to constrain the parameters $\overrightarrow{%
\lambda }\equiv (\epsilon _{a},$ $\tilde{\epsilon}_{a})$:
\begin{equation}
\chi ^{2}(\overrightarrow{\lambda })=\sum_{i}\frac{%
(R_{exp}^{i}-R_{pre}^{i}(1+\alpha ^{i}))^{2}}{(\sigma _{stat}^{i})^{2}}%
+\left( \frac{\alpha ^{i}}{\sigma _{\alpha }^{i}}\right) ^{2},
\label{eq:chi_ssq}
\end{equation}%
where $i$ runs over the solar neutrino sources \textit{pp}, $^{7}$Be and 
\textit{pep}. In Eq.\ (\ref{eq:chi_ssq}), $R_{exp}$\ are the experimental
event rates observed at Borexino in phase-I and phase-II with $\sigma _{stat}
$\ the statistical uncertainties for each of the five measurements, while 
$R_{pre}$\ is the predicted event rate corresponding to each experiment,
calculated using Eq.\ (\ref{eq:totrate}). The predicted and measured event
rates are quoted in Table \ref{tab:borex}, as well as our calculated values
for comparison. 
We take the neutrino energy window of 100-420 keV
for calculating the $pp$-neutrino event rate. The obtained results for the
SM case are given in Table \ref{tab:borex}. 

In Eq.\ (\ref{eq:chi_ssq}), we also add a penatly term corresponding to each
measurement to account for the theoretical uncertainities in the solar flux
model for the three solar spectra and from the oscillation parameters,
mostly from $\theta _{12}$ since $\theta _{13}$ and $\Delta m_{12}^{2}$ are
known very well. In Table \ref{tab:borex} we quote the percentage
uncertainties for each spectrum using Borexino's predicted event rates. We
use the predicted percentage uncertainties as the constraints ($\sigma
_{\alpha }$) on the pull parameters ($\alpha ^{i}$). Since the five measured
event rate values given in Table \ref{tab:borex} are already
background-subtracted we do not include any background terms in our $\chi
^{2}$-model. Additionally, since we are working with the event rates  we are less affected by 
details of the detector energy resolutions or detector
response, etc. 

As stated earlier, since we are using a simple\textbf{\ }$\chi ^{2}$-model that is based on the total event rate analysis corresponding to each low energy component of the solar spectrum, we consider only the statistical uncertainties and do not take into account the experimental correlated or uncorrelated systematic errors. 
The statistical analysis we have implemented here has already been used for
phenomenological new physics studies in Refs.\ 
\cite{Deniz:2010mp,Khan:2017oxw} and others. The validity of the $%
\chi ^{2}$-model used\ here has been cross-checked for estimating the
neutrino magnetic moments for the same data in Ref.\ 
\cite{Khan:2017djo}. The results of that work are in good agreement with
those obtained by Borexino in Ref.\  \cite%
{Borexino:2017fbd} for phase-II data. As an explicit comparison, the analysis from Ref.\ \cite{Khan:2017oxw} applied to phase-II data without CNO data gives for the weak mixing angle $\sin^2 \theta_W = 0.229 \pm 0.038$, to be compared to Borexino's result \cite{Agarwalla:2019smc} of $\sin^2 \theta_W = 0.229 \pm 0.026$. 
The limits we will present in what follows are therefore conservative. 

We emphasize that full agreement between our results and Borexino's result of the vector and axial vector parameters \cite{Agarwalla:2019smc} cannot be anticipated since we are not including the CNO data in our analysis as the direct rate measurement by Borexino is not available, we rather use data from phase I and phase II while Borexino uses only phase II data. Also, since the $\beta$ spectrum of $^{210}$Bi and $^{85}$Kr as backgrounds coincide with $^{7}$Be spectrum that might lead to these mild modifications. Spectral analysis of all the general parameters using full set of the Borexino's data is beyond the scope of this work. In the analysis we use the values of masses and mixing angles from PDG \cite{Tanabashi:2018oca}. 

\section{\label{sec:res}Results}

\noindent Having described our fitting procedure, we present here the
results. As the produced electron neutrinos oscillate also to muon and tau
neutrinos, we study two scenarios: (i) new interactions appear only
for $\nu _{e}$, and (ii) new interactions appear only for $\nu _{\mu /\tau }$
(recall that we do not distinguish both flavors).

\begin{figure}[p]
\begin{center}
\includegraphics[width=5.5in]{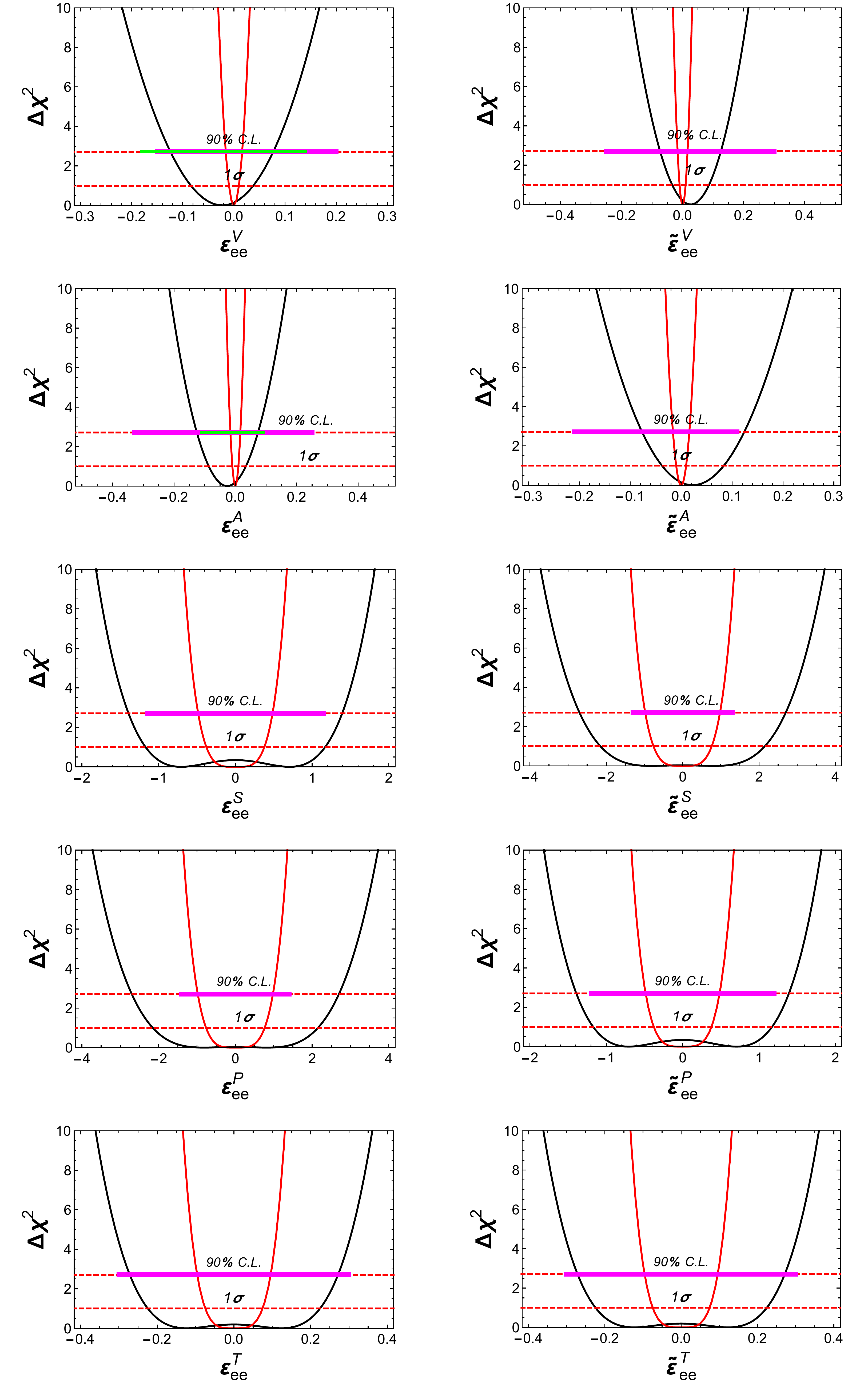}
\end{center}
\caption{Constraints on general neutrino interactions, see Eq.\ (\protect
\ref{eq:LBSM}), for electron neutrinos. The black line is the limit
obtained from Borexino event rates, the red line for hypothetical future
measurements with event rate precision of 1\%, see Sec.\  \protect \ref%
{sec:fut}. Indicated are the $1\protect \sigma $ and 90\% C.L.\ projections.
The thick horizontal magenta line is the limit obtained from TEXONO  data,
taken from \protect \cite{Rodejohann:2017vup}. The green horizontal lines at 90\% C.L.\ projections
in the top two left figures are bounds from ref. \cite{Agarwalla:2019smc}.}
\label{fig:eps_e}
\end{figure}

\begin{figure}[p]
\begin{center}
\includegraphics[width=5.5in]{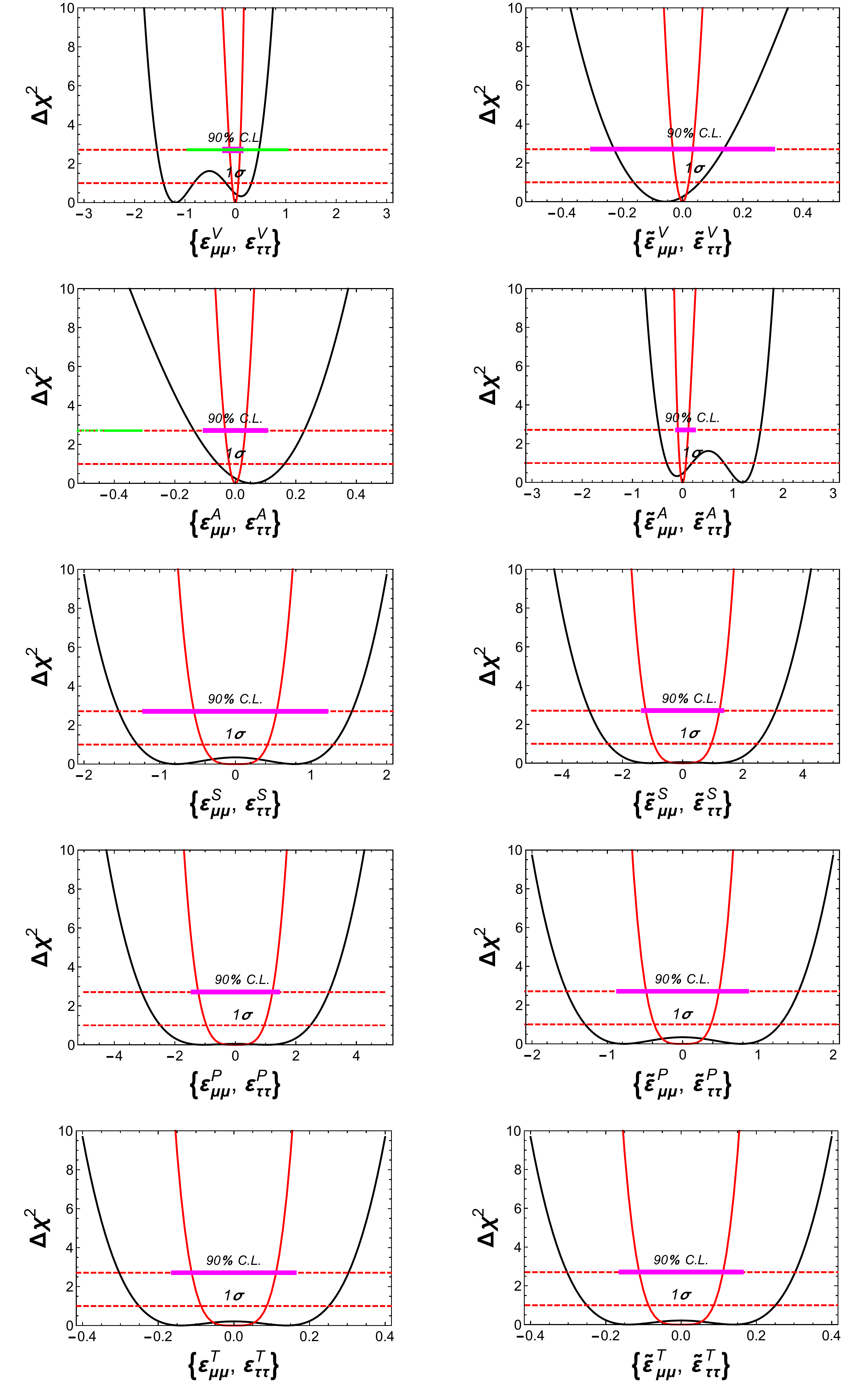}
\end{center}
\caption{Constraints on general neutrino interactions, see Eq.\ (\protect
\ref{eq:LBSM}), for muon/tau  neutrinos. The black line is the limit
obtained from Borexino event rates, the red line for hypothetical future
measurements with event rate precision of 1\%, see Sec.\  \protect \ref%
{sec:fut}. Indicated are the $1\protect \sigma $ and 90\% C.L.\ projections.
The thick horizontal magenta line is the limit on muon neutrino general
interactions obtained from CHARM-II data, taken from \protect \cite%
{Rodejohann:2017vup}. The green horizontal lines at 90\% C.L.\ projections
in the top two left figures are bounds from ref. \cite{Agarwalla:2019smc}. The dots on the left of the green line in the 2nd graph shows that the lower bound goes up to -0.72 which is beyond the chosen scale of our analysis.}
\label{fig:eps_mt}
\end{figure}

Fig.\  \ref{fig:eps_e} shows the result of our $\chi^2$-fit for the general interactions of
electron neutrinos. The constraints are compared to previous
constraints obtained in Ref.\  \cite{Rodejohann:2017vup} using TEXONO reactor anti-neutrino data.
Borexino improves the limits on $\epsilon_{ee}^{V}$, $\tilde{\epsilon}%
_{ee}^{V}$, $\epsilon_{ee}^{A}$, $\epsilon_{ee}^{T}$ and $\tilde{\epsilon}%
_{ee}^{T}$. 
Fig.\  \ref{fig:eps_mt} shows the fit result for new physics acting only on
muon/tau neutrinos. The constraints are compared to a previous
constraint obtained in Ref.\  \cite{Rodejohann:2017vup} using CHARM-II data.
Borexino improves the limit on $\tilde{\epsilon}_{\mu \mu }^{V}$. 
For $\epsilon _{\tau \tau }^{a}$ and $\tilde{%
\epsilon}_{\tau \tau }^{a}$ these are the very first limits. The
numerical values of the constraints in Figs.\  \ref{fig:eps_e} and \ref%
{fig:eps_mt} are given in Tabs.\  \ref{tab:eps_e} and \ref{tab:eps_mt}. 

\begin{table*}[th]
\begin{center}
\begin{tabular}{c|c|c|c|c|c}
\hline \hline
parameter & this study (solar) & ref. \cite{Agarwalla:2019smc} & ref. \cite{Bolanos:2008km} & ref. \cite{Rodejohann:2017vup, Khan:2014zwa, Khan:2016uon} & 
future (solar) \\ \hline
$\varepsilon _{ee}^{V}$ & \multicolumn{1}{|l|}{$[-0.12,0.08]$} & $%
[-0.18,0.14]$ & $[-0.31,0.65]$ & \multicolumn{1}{|l|}{$[-0.13,0.20]$} & 
\multicolumn{1}{|l}{$[-0.016,\ 0.016]$} \\ \hline
$\varepsilon _{ee}^{A}$ & \multicolumn{1}{|l|}{$[-0.13,0.07]$} & $%
[-0.11,0.08]$ & $[-0.23,0.53]$ & \multicolumn{1}{|l|}{$[-0.32,0.22]$} & 
\multicolumn{1}{|l}{$[-0.016,0.016]$} \\ \hline
$\varepsilon _{ee}^{S}$ & \multicolumn{1}{|l|}{$[-1.4,1.4]$} & $-$ & $-$ & 
\multicolumn{1}{|l|}{$[-1.1,1.1]$} & \multicolumn{1}{|l}{$[-0.49,0.49]$} \\ 
\hline
$\varepsilon _{ee}^{P}$ & \multicolumn{1}{|l|}{$[-2.7,\ 2.7]$} & $-$ & $-$ & 
\multicolumn{1}{|l|}{$[-1.3,1.3]$} & \multicolumn{1}{|l}{$[-0.98,~0.98]$} \\ 
\hline
$\varepsilon _{ee}^{T}$ & \multicolumn{1}{|l|}{$[-0.27,0.27]$} & $-$ & $-$ & 
\multicolumn{1}{|l|}{$[-0.31,0.30]$} & \multicolumn{1}{|l}{$[-0.09,0.09]$}
\\ \hline
$\widetilde{\varepsilon }_{ee}^{V}$ & \multicolumn{1}{|l|}{$[-0.07,0.13]$} & 
$-$ & $-$ & \multicolumn{1}{|l|}{$[-0.23,0.29]$} & \multicolumn{1}{|l}{$%
[-0.016,0.016]$} \\ \hline
$\widetilde{\varepsilon }_{ee}^{A}$ & \multicolumn{1}{|l|}{$[-0.08,0.13]$} & 
$-$ & $-$ & \multicolumn{1}{|l|}{$[-0.21,0.11]$} & \multicolumn{1}{|l}{$%
[-0.016,0.016]$} \\ \hline
$\widetilde{\varepsilon }_{ee}^{S}$ & \multicolumn{1}{|l|}{$[-2.7,~2.7]$} & $%
-$ & $-$ & \multicolumn{1}{|l|}{$[-1.3,1.3]$} & \multicolumn{1}{|l}{$%
[-0.98,0.98]$} \\ \hline
$\widetilde{\varepsilon }_{ee}^{P}$ & \multicolumn{1}{|l|}{$[-1.4,1.4]$} & $-
$ & $-$ & \multicolumn{1}{|l|}{$[-1.1,1.1]$} & \multicolumn{1}{|l}{$%
[-0.48,0.48]$} \\ \hline
$\widetilde{\varepsilon }_{ee}^{T}$ & \multicolumn{1}{|l|}{$[-0.27,0.27]$} & 
$-$ & $-$ & \multicolumn{1}{|l|}{$[-0.31,0.29]$} & \multicolumn{1}{|l}{$%
[-0.09,0.09]$} \\ \hline \hline
\end{tabular}%
\end{center}
\caption{90\% C.L.\ constraints on general neutrino interactions for
electron neutrinos, see Eq.\ (\protect \ref{eq:LBSM}), obtained from Borexino
data, displayed in Fig. 1. Given also are previous limits and hypothetical
future constraints. In 4th and 5th column the bounds were translated from the corresponding references by the relation $\epsilon_{\alpha\alpha}^{V,A} = \epsilon_{\alpha\alpha}^{R}\pm \epsilon_{\alpha\alpha}^{L}$.}
\label{tab:eps_e}
\end{table*}

\begin{table*}[th]
\begin{center}
\begin{tabular}{c|c|c|c|c|c}
\hline \hline
parameter & this study (solar) & ref. \cite{Agarwalla:2019smc} & ref. \cite{Bolanos:2008km} & ref. \cite{Rodejohann:2017vup, Khan:2014zwa, Khan:2016uon} & 
future (solar) \\ \hline
$\varepsilon _{\mu \mu }^{V}/\varepsilon _{\tau \tau }^{V}$ & 
\multicolumn{1}{|l|}{$[-1.5,0.5]$} & $[-0.94,1.03]$ & $[-1.21,0.42]$ & 
\multicolumn{1}{|l|}{$[-0.22,0.08]$} & \multicolumn{1}{|l}{$[-0.1,\ 0.1]\ $}
\\ \hline
$\varepsilon _{\mu \mu }^{A}/\varepsilon _{\tau \tau }^{A}$ & 
\multicolumn{1}{|l|}{$[-0.14,0.23]$} & $[-0.72,-0.31]$ & $[-0.89,0.20]$ & 
\multicolumn{1}{|l|}{$[-0.08,0.08]$} & \multicolumn{1}{|l}{$[-0.03,0.03]$}
\\ \hline
$\varepsilon _{\mu \mu }^{S}/\varepsilon _{\tau \tau }^{S}$ & 
\multicolumn{1}{|l|}{$[-1.5,1.5]$} & $-$ & $-$ & \multicolumn{1}{|l|}{$%
[-0.83,0.83]$} & \multicolumn{1}{|l}{$[-0.5,0.5]$} \\ \hline
$\varepsilon _{\mu \mu }^{P}/\varepsilon _{\tau \tau }^{P}$ & 
\multicolumn{1}{|l|}{$[-3,\ 3]$} & $-$ & $-$ & \multicolumn{1}{|l|}{$%
[-0.83,0.83]$} & \multicolumn{1}{|l}{$[-1.22,~1.22]$} \\ \hline
$\varepsilon _{\mu \mu }^{T}/\varepsilon _{\tau \tau }^{T}$ & 
\multicolumn{1}{|l|}{$[-0.271,0.271]$} & $-$ & $-$ & \multicolumn{1}{|l|}{$%
[-0.15,0.15]$} & \multicolumn{1}{|l}{$[-0.1,0.1]$} \\ \hline
$\widetilde{\varepsilon }_{\mu \mu }^{V}/\widetilde{\varepsilon }_{\tau \tau
}^{V}$ & \multicolumn{1}{|l|}{$[-0.4,0.8]$} & $-$ & $-$ & 
\multicolumn{1}{|l|}{$[-0.09,0.08]$} & \multicolumn{1}{|l}{$[-0.04,0.04]$}
\\ \hline
$\widetilde{\varepsilon }_{\mu \mu }^{A}/\widetilde{\varepsilon }_{\tau \tau
}^{A}$ & \multicolumn{1}{|l|}{$[-1.2,0.9]$} & $-$ & $-$ & 
\multicolumn{1}{|l|}{$[-0.09,0.22]$} & \multicolumn{1}{|l}{$[-0.1,0.1]$} \\ 
\hline
$\widetilde{\varepsilon }_{\mu \mu }^{S}/\widetilde{\varepsilon }_{\tau \tau
}^{S}$ & \multicolumn{1}{|l|}{$[-3,~3]$} & $-$ & $-$ & \multicolumn{1}{|l|}{$%
[-0.83,0.83]$} & \multicolumn{1}{|l}{$[-1.2,1.2]$} \\ \hline
$\widetilde{\varepsilon }_{\mu \mu }^{P}/\widetilde{\varepsilon }_{\tau \tau
}^{P}$ & \multicolumn{1}{|l|}{$[-1.54,1.54]$} & $-$ & $-$ & 
\multicolumn{1}{|l|}{$[-0.83,0.83]$} & \multicolumn{1}{|l}{$[-0.54,0.54]$}
\\ \hline
$\widetilde{\varepsilon }_{\mu \mu }^{T}/\widetilde{\varepsilon }_{\tau \tau
}^{T}$ & \multicolumn{1}{|l|}{$[-0.3,0.3]$} & $-$ & $-$ & 
\multicolumn{1}{|l|}{$[-0.15,0.15]$} & \multicolumn{1}{|l}{$[-0.1,0.1]$} \\ 
\hline \hline
\end{tabular}%
\end{center}
\caption{90\% C.L.\ constraints on general neutrino interactions for
muon/tau neutrinos, see Eq.\ (\protect \ref{eq:LBSM}), obtained from Borexino
data, displayed in Fig. 2. Given also are previous limits (on muon
neutrinos) and hypothetical future constraints. In 4th and 5th column the bounds were translated from the corresponding references by the relation $\epsilon_{\alpha\alpha}^{V,A} = \epsilon_{\alpha\alpha}^{R}\pm \epsilon_{\alpha\alpha}^{L}$.}
\label{tab:eps_mt}
\end{table*}

\begin{figure}[th]
\begin{center}
\includegraphics[width=6in]{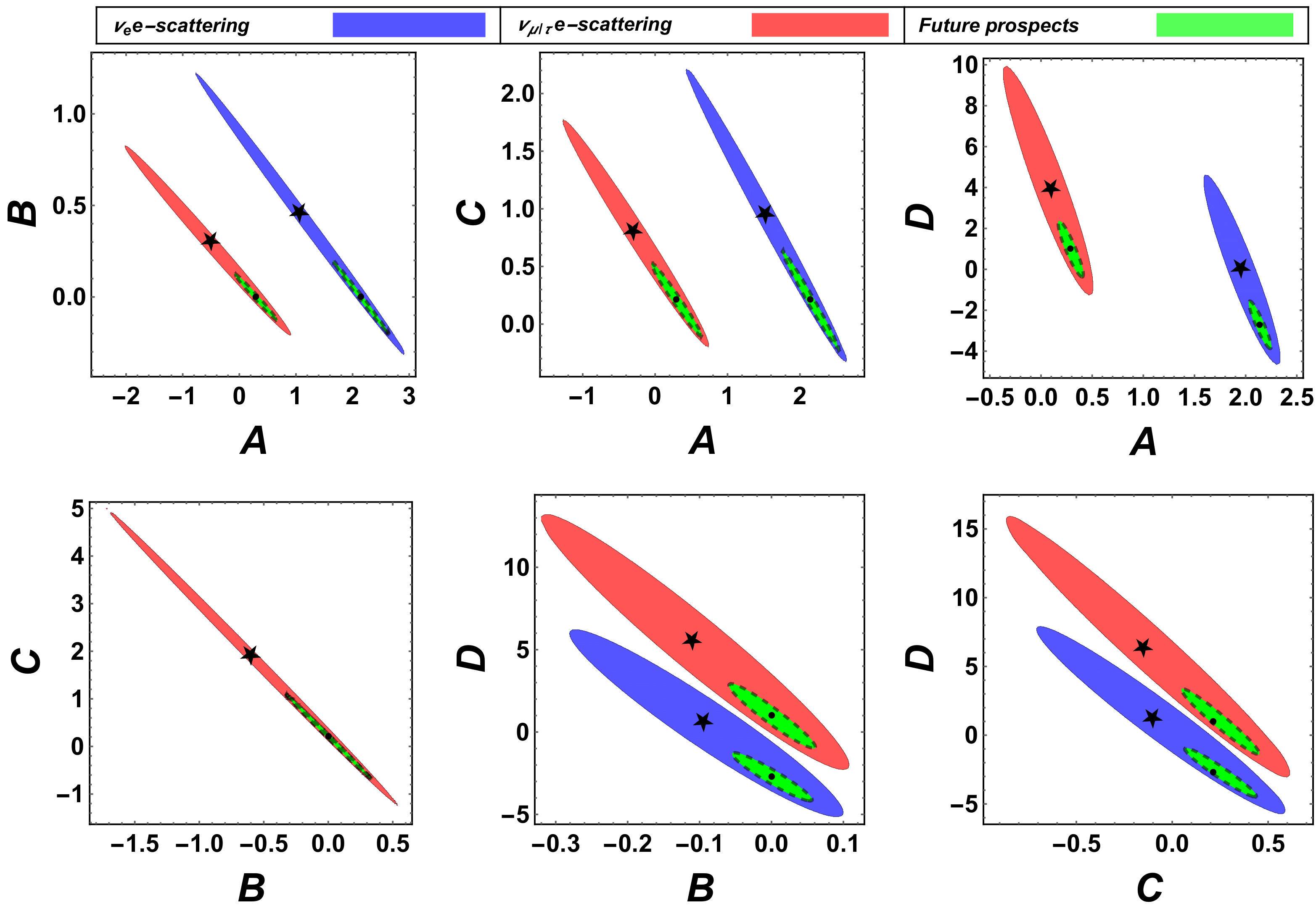}
\end{center}
\caption{90 \% C.L.\ constraints on cross section parameters $A$, $B$, $C$
and $D$ that appear in the total cross section Eq.\ (\protect \ref{eq:xsABCD}%
). The Standard Model values are indicated as black dots, our best-fit values 
are the black stars. The yellow region assumes a hypothetical future
measurement with 1\% precision, see Sec.\  \protect \ref{sec:fut}. }
\label{fig:ABCD}
\end{figure}

It is also useful to give constraints on the parameters $A$, $B$, $C$ and $D$ 
that appear in the total cross section Eq.\ (\ref{eq:xsABCD}). The result is 
given in Fig.\  \ref{fig:ABCD} and Tab.\  \ref{tab:ABCD} for electron and
muon/tau  neutrinos. We have performed here two-parameter fits setting the other two parameters to their SM 
values. The SM values of these parameters are given in the last two columns of 
Tab.\  \ref{tab:ABCD}, which in particular is $B = 0$. For future experiments (see Sec.\ \ref{sec:fut}) we assumed the SM values of $A,B,C$ and $D$. 

\begin{table*}[ht]
\begin{center}
\begin{tabular}{l|l|l|l|l|l|l}
\hline \hline
parameter & bounds $(\nu _{e}e)$ & bounds ($\nu _{\mu /\tau }e)$ & best-fit (%
$\nu _{e}e)$ & best-fit $(\nu _{\mu /\tau }e)$ & SM$\ (\nu _{e}e)$ & SM$\
(\nu _{\mu /\tau }e)$ \\ \hline
$A_{\alpha e}$ & $[2,2.3]$ & $[0.15,0.45]$ & $2.2$ & $0.31$ & $2.12$ & $0.29$
\\ \hline
$B_{\alpha e}$ & $[-0.05,0.07]$ & $[-0.05,0.07]$ & $0.01$ & $0.01$ & $0$ & $%
0 $ \\ \hline
$C_{\alpha e}$ & $[0.11,0.41]$ & $[0.11,0.41]$ & $0.24$ & $0.24$ & $0.21$ & $%
0.21$ \\ \hline
$D_{\alpha e}$ & $[-3.5,-0.2]$ & $[-0.2,3.5]$ & $-2.3$ & $1.5$ & $-2.7$ & $%
0.99$ \\ \hline \hline
\end{tabular}%
\end{center}
\caption{90\% C.L.\ constraints on the cross section parameters $A$, $B$, $C$
and $D$ that appear in the total cross section Eq.\ (\protect \ref{eq:xsABCD}%
) corresponding to Fig. 3. Dirac neutrinos are assumed, and the Standard Model values for the
parameters are also given. }
\label{tab:ABCD}
\end{table*}

\subsection{Comparison with the conventional NSI and other studies}
We note that recently BOREXINO collaboration has published constraints on non-standard interactions 
$\epsilon_{\alpha\alpha}^{L,R}$ using phase-II data only \cite{Agarwalla:2019smc}. Using the general relation $\epsilon_{\alpha\alpha}^{V,A} = \epsilon_{\alpha\alpha}^{R}\pm \epsilon_{\alpha\alpha}^{L}$, we translate their bounds in our notation and present them in the 3rd and 4th column of Table II and III for comparison. The
difference between the two results can be understood as follows. We use rates of five BOREXINO experiments from both phase I and II while ref. [38] uses only the phase II data. We use a different statistical model for parameter fitting than them. They divide the pull term in their statistical function by a factor of 2. We have assumed the maximal mixing assumption for $\theta_{23}$ while ref. \cite{Agarwalla:2019smc} has taken the opposite. An exact comparison would be difficult however given the above differences in approaches and data sets, the agreement is reasonable.

We note that the comparison to previous limits from 
reactor anti-neutrinos can be partly understood from the form of the
cross-section in Eq.\ (\ref{diffcros}). The term proportional to $C$ is suppressed by about a factor of $3$ in the total cross-section with respect to $A$ in the energy regime of BOREXINO measurements considered in this work. For anti-neutrinos $A$ and $C$ are interchanged in the cross section, which means that the solar data is more sensitive to the left-handed
neutral currents ($V-A$), that is, to parameters in the definition of $A$ here, 
while the reactor data is more sensitive to the right-handed parameters ($V+A$), occurring here in the definition of $C$ in Eq.\ (6). This argument is valid only for $\nu _{e}e$-scattering.
For the $\nu _{\mu /\tau }e$-scatterings, the bounds here are
comparable to those from CHARM-II which is sensitive to both left-handed and
right-handed couplings. However, as we can see that from Tabs.\  \ref{tab:eps_e}
and \ref{tab:eps_mt}, solar data give better constraints on tensor
interactions while weaker or comparable constraints on scalar and
pseudo-scalar interactions.

We note that a recent work \cite{Ge:2018uhz} introduced scalar NSI, similar to $a=S$ in our Eq.~(4), to study the matter effect in neutrino oscillation. The strength considered in Ref.\ \cite{Ge:2018uhz} corresponds to $\epsilon^{S} \sim 10^{6}$ to $10^{9}$. Such strong interactions are not compatible with most neutrino scattering data unless they are mediated by very light bosons. The scenario of light mediators has been further investigated in Ref.\ \cite{Smirnov:2019cae}, which concludes that scalar NSI cannot cause significant effects in current experiments. In our work, we do not consider light mediators and always focus on four-fermion effective operators. Therefore, for scalar interactions not much stronger than the SM interactions (i.e., $\epsilon^{S}\leq{\cal O}(1)$), the matter effect is negligible.

\subsection{ Future Prospects from solar Data \label{sec:fut}}

There are several ideas floating around to further improve the precision on
solar neutrino measurements with a precision of 1\% or better. The main
motivations behind these projects are the determination of the correct
metallicity (low or high) solar model and photon fluxes from the Sun, a more
stringent test of the LMA-MSW solution of the neutrinos propagating through
the solar matter and to explore exotic properties related to the solar
neutrinos. One such project is the Jinping experiment \cite%
{JinpingNeutrinoExperimentgroup:2016nol}. In addition, future large scale
dark matter direct detection experiments can provide precise solar neutrino
measurements \cite{Baudis:2013qla,Cerdeno:2016sfi,Aalbers:2016jon}, ideas to
use future long-baseline neutrino oscillation far detectors as solar
neutrino experiments are also present \cite{Capozzi:2018dat}.

As for different types of potential future experiments the precision of the
individual solar neutrino sources is different, we conservatively adopt for
simplicity that all the three low energy solar neutrino fluxes ($pp$, $pep$ and $^{7}$Be) will have been 
measured with a 1\% precision and assume the SM values of the various $\epsilon $, $\tilde{%
\epsilon}$ or $A,B,C,D$. With this projected precision we simulate our data for the same type and size of Borexino detector and then fit all the parameters
in a similar fashion as was done for the real data. The results of this analysis are displayed with red color distributions in Figs.\ \ref{fig:eps_e} and Fig.\  \ref{fig:eps_mt} and with green color ellipses in Fig.\  \ref{fig:ABCD}. 

As clear from Figs.\ \ref{fig:eps_e} and Fig.\  \ref{fig:eps_mt}, and can
be read off from Tabs.\  \ref{tab:eps_e} and \ref{tab:eps_mt}, the future 
solar data with 1\% precision will improve the current bounds on non-standard
vector, axial-vector and tensor interactions by more than one order of 
magnitude while for the scalar and pseudo-scalar ones they will improve by a
factor of 3 to 5, in general. The future constraints at 90\% C.L.\ on the
parameters $A,B,C$ and $D$ are also shown with green color 
ellipses overlaid on the current constraints for comparison in 
Fig.\  \ref{fig:ABCD}.

\section{The difference between Dirac and Majorana neutrinos \label{sec:diffDM}}
\noindent
Throughout this paper, our analyses assumed 
Dirac neutrinos. As we have previously mentioned, for Majorana neutrinos 
some interactions are absent because they have fewer
degrees of freedom than Dirac neutrinos. More explicitly, a Dirac neutrino
spinor consists of both left-handed and right-handed components ($\nu_{L}$
and $\nu_{R}$):
\begin{equation}
\nu_{D}=\nu_{L}+\nu_{R}\,,\label{eq:DLR}
\end{equation}
where $\nu_{L}$ and $\nu_{R}$ are two independent fermionic degrees
of freedom. A Majorana neutrino spinor is conventionally defined as
\begin{equation}
\nu_{M}=\nu_{L}+\nu_{L}^{c}\,,\label{eq:MLR}
\end{equation}
so that $\nu_{M}=\nu_{M}^{c}$. Here $\nu_{L}^{c}$ is the charge
conjugate of $\nu_{L}$, containing essentially the same Weyl spinor
as $\nu_{L}$. Expanding $\overline{\nu}\Gamma\nu$ in terms of the
chiral components, one can immediately see that some interactions cannot
exist for $\nu_{M}$. For instance,
\begin{eqnarray}
\overline{\nu_{D}}\gamma^{\mu}\nu_{D} & = & \overline{\nu_{L}}\gamma^{\mu}\nu_{L}+\overline{\nu_{R}}\gamma^{\mu}\nu_{R}\,,\label{eq:x}\\
\overline{\nu_{M}}\gamma^{\mu}\nu_{M} & = & \overline{\nu_{L}}\gamma^{\mu}\nu_{L}+\overline{\nu_{L}^{c}}\gamma^{\mu}\nu_{L}^{c}=\overline{\nu_{L}}\gamma^{\mu}\nu_{L}-\overline{\nu_{L}}\gamma^{\mu}\nu_{L}=0\,,\label{eq:x-1}
\end{eqnarray}
which implies that {\it flavor-diagonal} vector interactions cannot
exist for Majorana neutrinos.  Likewise, one can check that for the
tensor interactions 
\begin{eqnarray}
\overline{\nu_{D}}\sigma^{\mu\nu}\nu_{D} & = & \overline{\nu_{R}}\sigma^{\mu\nu}\nu_{L}+\overline{\nu_{L}}\sigma^{\mu\nu}\nu_{R}\,,\label{eq:x-2}\\
\overline{\nu_{M}}\sigma^{\mu\nu}\nu_{M} & = & \overline{\nu_{L}^{c}}\sigma^{\mu\nu}\nu_{L}+\overline{\nu_{L}}\sigma^{\mu\nu}\nu_{L}^{c}=0\,.\label{eq:x-3}
\end{eqnarray}
\begin{table}
\centering
\begin{tabular}{l|l|l||l|l}
\hline \hline 
$\overline{\nu_{\alpha}}\Gamma\nu_{\beta}$ & Dirac: $\nu_{D}=\nu_{L}+\nu_{R}$ & Dirac: $\nu_{D}=\left(\begin{array}{c}
\chi\\
\overline{\xi}
\end{array}\right)$ & Majorana: $\nu_{M}=\nu_{L}+\nu_{L}^{c}$ & Majorana: $\nu_{M}=\left(\begin{array}{c}
\chi\\
\overline{\chi}
\end{array}\right)$\tabularnewline
\hline 
$\overline{\nu_{\alpha}}\nu_{\beta}$ & $\overline{\nu_{\alpha R}}\nu_{\beta L}+\overline{\nu_{\alpha L}}\nu_{\beta R}$ & $\mbox{\ensuremath{\xi}}_{\alpha}\chi_{\beta}+\overline{\mbox{\ensuremath{\chi}}_{\alpha}}\overline{\xi_{\beta}}$ & $\overline{\nu_{\alpha L}^{c}}\nu_{\beta L}+\overline{\nu_{\alpha L}}\nu_{\beta L}^{c}$ & $\mbox{\ensuremath{\chi}}_{\alpha}\chi_{\beta}+\overline{\mbox{\ensuremath{\chi}}_{\alpha}}\overline{\chi_{\beta}}$\tabularnewline
\hline 
$\overline{\nu_{\alpha}}i\gamma^{5}\nu_{\beta}$ & $-i\overline{\nu_{\alpha R}}\nu_{\beta L}+i\overline{\nu_{\alpha L}}\nu_{\beta R}$ & $-i\mbox{\ensuremath{\xi}}_{\alpha}\chi_{\beta}+i\overline{\mbox{\ensuremath{\chi}}_{\alpha}}\overline{\xi_{\beta}}$ & $-i\overline{\nu_{\alpha L}^{c}}\nu_{\beta L}+i\overline{\nu_{\alpha L}}\nu_{\beta L}^{c}$ & $-i\mbox{\ensuremath{\chi}}_{\alpha}\chi_{\beta}+i\overline{\mbox{\ensuremath{\chi}}_{\alpha}}\overline{\chi_{\beta}}$\tabularnewline
\hline 
$\overline{\nu_{\alpha}}\gamma^{\mu}\nu_{\beta}$ & $\overline{\nu_{\alpha L}}\gamma^{\mu}\nu_{\beta L}+\overline{\nu_{\alpha R}}\gamma^{\mu}\nu_{\beta R}$ & $\overline{\chi_{\alpha}}\overline{\sigma^{\mu}}\chi_{\beta}+\xi_{\alpha}\sigma^{\mu}\overline{\xi_{\beta}}$ & $\overline{\nu_{\alpha L}}\gamma^{\mu}\nu_{\beta L}-\overline{\nu_{\beta L}}\gamma^{\mu}\nu_{\alpha L}$
 & $\overline{\chi_{\alpha}}\overline{\sigma^{\mu}}\chi_{\beta}+\chi_{\alpha}\sigma^{\mu}\overline{\chi_{\beta}}$\tabularnewline
\hline 
$\overline{\nu_{\alpha}}\gamma^{\mu}\gamma^{5}\nu_{\beta}$ & $-\overline{\nu_{\alpha L}}\gamma^{\mu}\nu_{\beta L}+\overline{\nu_{\alpha R}}\gamma^{\mu}\nu_{\beta R}$ & $-\overline{\chi_{\alpha}}\overline{\sigma^{\mu}}\chi_{\beta}+\xi_{\alpha}\sigma^{\mu}\overline{\xi_{\beta}}$ & $-\overline{\nu_{\alpha L}}\gamma^{\mu}\nu_{\beta L}-\overline{\nu_{\beta L}}\gamma^{\mu}\nu_{\alpha L}$ & $-\overline{\chi_{\alpha}}\overline{\sigma^{\mu}}\chi_{\beta}+\chi_{\alpha}\sigma^{\mu}\overline{\chi_{\beta}}$\tabularnewline
\hline 
$\overline{\nu_{\alpha}}\sigma^{\mu\nu}\nu_{\beta}$ & $\overline{\nu_{\alpha R}}\sigma^{\mu\nu}\nu_{\beta L}+\overline{\nu_{\alpha L}}\sigma^{\mu\nu}\nu_{\beta R}$ & $\xi_{\alpha}\sigma_{2\times2}^{\mu\nu}\chi_{\beta}+\overline{\chi_{\alpha}}\overline{\sigma_{2\times2}^{\mu\nu}}\overline{\xi_{\beta}}$ & $\overline{\nu_{\alpha L}^{c}}\sigma^{\mu\nu}\nu_{\beta L}+\overline{\nu_{\alpha L}}\sigma^{\mu\nu}\nu_{\beta L}^{c}$
 & $\chi_{\alpha}\sigma_{2\times2}^{\mu\nu}\chi_{\beta}+\overline{\chi_{\alpha}}\overline{\sigma_{2\times2}^{\mu\nu}}\overline{\chi_{\beta}}$\tabularnewline
\hline \hline 
\end{tabular}\caption{Neutrino spinor products ($\overline{\nu_{\alpha}}\Gamma\nu_{\beta}$)
written explicitly in terms of the chiral components (2nd and
4th columns), or in terms of the Weyl spinors (3rd and 5th columns).
Here $\sigma^{\mu}=(1,\ \vec{\sigma})$ and $\overline{\sigma^{\mu}}=(1,\ -\vec{\sigma})$
are the Lorentz-covariant Pauli matrices, $\sigma_{2\times2}^{\mu\nu}$
and $\overline{\sigma_{2\times2}^{\mu\nu}}$ are defined as $\sigma_{2\times2}^{\mu\nu}\equiv\frac{i}{2}\left(\sigma^{\mu}\overline{\sigma^{\nu}}-\sigma^{\nu}\overline{\sigma^{\mu}}\right)$
and $\overline{\sigma_{2\times2}^{\mu\nu}}\equiv\frac{i}{2}\left(\overline{\sigma^{\mu}}\sigma^{\nu}-\overline{\sigma^{\nu}}\sigma^{\mu}\right)$.
 \label{tab:DM_chiral}}
\end{table}
\noindent However, {\it off-diagonal} vector and tensor interactions are possible for Majorana neutrinos. This is in analogy to the well-known
fact that Majorana neutrinos cannot have flavor
diagonal magnetic moments (which couple photons to neutrinos via $\sigma^{\mu\nu}$)
but can have flavor transition magnetic moments -- see, e.g., \cite{Xu:2019dxe}. In Tab.~\ref{tab:DM_chiral},
we list the chiral expansion for all possible products with flavor
indices included. The results for Majorana neutrinos can be simply
obtained by replacing $\nu_{R}$ in the Dirac column with $\nu_{L}^{c}$. 

Let us inspect the symmetry of these products when the flavor indices
$\alpha$ and $\beta$ are interchanged.  For any two general spinors
$\psi_{\alpha}$ and $\psi_{\beta}$ (applicable to both Dirac and
Majorana), it can be verified that $\overline{\psi_{\alpha}^{c}}\Gamma\psi_{\beta}$
is symmetric with respect to $\alpha\leftrightarrow\beta$ for $\Gamma=(1,\ i\gamma^{5},\ \gamma^{\mu}\gamma^{5})$,
and becomes anti-symmetric for $\Gamma=(\gamma^{\mu},\ \sigma^{\mu\nu},\ \sigma^{\mu\nu}\gamma^{5})$,
 e.g., $\overline{\psi_{\alpha}^{c}}\psi_{\beta}=\overline{\psi_{\beta}^{c}}\psi_{\alpha}$,
$\overline{\psi_{\alpha}^{c}}\gamma^{\mu}\psi_{\beta}=-\overline{\psi_{\beta}^{c}}\gamma^{\mu}\psi_{\alpha}$
and $\overline{\psi_{\alpha}^{c}}\sigma^{\mu\nu}\psi_{\beta}=-\overline{\psi_{\beta}^{c}}\sigma^{\mu\nu}\psi_{\alpha}$.
Now for Majorana neutrinos, due to their self-conjugate property ($\nu_{M}=\nu_{M}^{c}$),
we have 
\begin{equation}
\overline{\nu_{M\alpha}}\Gamma\nu_{M\beta}=\overline{\nu_{M\alpha}^{c}}\Gamma\nu_{M\beta}=-\overline{\nu_{M\beta}^{c}}\Gamma\nu_{M\alpha}=-\overline{\nu_{M\beta}}\Gamma\nu_{M\alpha}~ {\rm for}~ \Gamma=(\gamma^{\mu},\ \sigma^{\mu\nu},\ \sigma^{\mu\nu}\gamma^{5})\,,\label{eq:x-4}
\end{equation}
and likewise
\begin{equation}
\overline{\nu_{M\alpha}}\Gamma\nu_{M\beta}=\overline{\nu_{M\beta}}\Gamma\nu_{M\alpha}~ {\rm for}~ \Gamma=(1,\ i\gamma^{5},\ \gamma^{\mu}\gamma^{5})\,.\label{eq:x-5}
\end{equation}
This implies that for Majorana neutrinos, the vector and tensor interactions
are flavor anti-symmetric; the scalar, pseudoscalar and axialvector
interactions are flavor symmetric. Therefore, 
\begin{equation}
{\rm Majorana:}\ \begin{cases}
\epsilon_{\alpha\beta}^{a}=-\epsilon_{\beta\alpha}^{a}\,,~ \tilde{\epsilon}_{\alpha\beta}^{a}=-\tilde{\epsilon}_{\beta\alpha}^{a},\  & ({\rm for}\ a=V,\ T)\\
\epsilon_{\alpha\beta}^{a}=\epsilon_{\beta\alpha}^{a}\,,~ \tilde{\epsilon}_{\alpha\beta}^{a}=\tilde{\epsilon}_{\beta\alpha}^{a} & ({\rm for}\ a=S,\ P,\ A)
\end{cases}.\label{eq:x-6}
\end{equation}
Note that these $\epsilon$ and $\tilde{\epsilon}$ matrices should
also be hermitian -- see footnote~\ref{fn:hermitian}, so Eq.~(\ref{eq:x-6})
is equivalent to 
\begin{equation}
{\rm Majorana:}\ \begin{cases}
{\rm Re}\,\epsilon_{\alpha\beta}^{a}={\rm Re}\,\tilde{\epsilon}_{\alpha\beta}^{a}=0,\  & ({\rm for}\ a=V,\ T)\\
{\rm Im}\,\epsilon_{\alpha\beta}^{a}={\rm Im}\,\tilde{\epsilon}_{\alpha\beta}^{a}=0 & ({\rm for}\ a=S,\ P,\ A)
\end{cases},\label{eq:x-7}
\end{equation}
which means $\epsilon_{\alpha\beta}^{a}$ and $\tilde{\epsilon}_{\alpha\beta}^{a}$
are real symmetric matrices for $a=S$, $P$, and $A$, and imaginary
anti-symmetric matrices for $a=V$, and $T$. In particular, the diagonal
parts of the vector and tensor coupling matrices should vanish, $\epsilon_{\alpha\alpha}^{V}=\tilde{\epsilon}_{\alpha\alpha}^{V}=\epsilon_{\alpha\alpha}^{T}=\tilde{\epsilon}_{\alpha\alpha}^{T}=0$.\\

In summary, the difference between Dirac and Majorana neutrinos in
the framework of this paper is that the $\epsilon$ and $\tilde{\epsilon}$
matrices for Majorana neutrinos are further constrained by Eq.~(\ref{eq:x-6}),
or equivalently, Eq.~(\ref{eq:x-7}). Thus our results based on Dirac
neutrinos are readily applicable to Majorana neutrinos except that
some of the couplings, namely flavor-diagonal $\epsilon_V$ and $\epsilon_T$, as well as $\tilde \epsilon_V$ and $\tilde \epsilon_T$,  should be absent.

\section{\label{sec:concl}Summary and Conclusions}

\noindent We have discussed here the sensitivity of Borexino to general
neutrino interactions. Assuming the presence of additional scalar,
pseudoscalar, vector, axialvector or tensor interactions we have
investigated how Borexino's measurements of $pp$, $pep$ and $^7$Be neutrino
event rates constrain the dimensionless (i.e.\ normalized to the Fermi
constant) interaction strength of the new interactions. Several previous
limits from TEXONO and CHARM-II are improved for the electron and muon
sector, while first limits on tau sector interactions were set. Our limits
are summarized in Figs.\  \ref{fig:eps_e}, \ref{fig:eps_mt} as well as Tabs.\ %
\ref{tab:eps_e} and \ref{tab:eps_mt}. We focused on Dirac neutrinos, and
detailed the difference to Majorana neutrinos. Future prospects on the
limits were also considered. Interpreting the interaction strengths as due
to some new exchanged boson with coupling $g_X$ and mass $M_X$ implies that $%
\epsilon$ or $\tilde \epsilon$ is given approximately by $(g_X^2/M_X^2) /G_F$.
This means that current (future) solar neutrino experiments are sensitive to
new physics of weak (TeV) scale and beyond.

\begin{acknowledgments}
\noindent WR was supported by the DFG with grant RO 2516/7-1 in the
Heisenberg program. AK thanks for a NPC fellowship at Fermilab where part of this
work was done, and is financially supported by the Alexander von Humboldt foundation.
\end{acknowledgments}

\bibliographystyle{apsrev4-1}
\bibliography{borex}

\end{document}